\documentclass[preprint,amsmath,amssymb,aps,prd,showpacs]{revtex4-1}

\usepackage{epsfig}
\usepackage{dcolumn}
\usepackage{bm}
\usepackage{tikz}
\usepackage{graphicx, graphics, color}
\usepackage{dcolumn}
\usepackage{bm}
\usepackage{hyperref}
\usepackage[mathlines]{lineno}

\newcommand{\be}{\begin{equation}}
\newcommand{\ee}{\end{equation}}
\newcommand{\ben}{\begin{eqnarray}}
\newcommand{\een}{\end{eqnarray}}

\newcommand{\cO}{{\cal O}}

\newcommand{\cL}{{\cal L}}

\newcommand{\p}{\partial}
\newcommand{\na}{\nabla}

\newcommand{\tphi}{\tilde \phi}
\newcommand{\tpsi}{\tilde \psi}

\newcommand{\tG}{\tilde G}
\newcommand{\tH}{{\tilde H}}
\newcommand{\tg}{\tilde g}

\newcommand{\ep}{\epsilon}

\newcommand{\ga}{\gamma}

\newcommand{\tR}{{\tilde R}}

\newcommand{\tna}{\tilde \na}

\begin{document}

\title{Uniqueness of photon sphere for Reissner-Nordstr\"om electric-magnetic system}
\author{Marek Rogatko} 
\email{rogat@kft.umcs.lublin.pl}
\affiliation{Institute of Physics, 
Maria Curie-Sklodowska University, 
20-031 Lublin, pl.~Marii Curie-Sklodowskiej 1, Poland}

\date{\today}

\begin{abstract}
Uniqueness of static, asymptotically flat, non-extremal {\it photon sphere} in Einstein-Maxwell spacetime with electric and magnetic charges has been proved.
Using conformal positive energy theorem, as well as, the positive mass theorem and adequate conformal transformations, we envisage 
the two alternative ways of proving that the exterior region
of a certain radius of the studied static {\it photon sphere}, is characterized by ADM mass, electric and magnetic charges. 
\end{abstract}

\maketitle

\section{Introduction}
In view of the first-ever obtained images of M87 and Milky Way supermassive black holes \cite{eht1,eht2}
and measure of
polarisation (a signature of magnetic fields close to the edge of a black hole) performed by the Event Horizon Telescope  (EHT) Collaboration
\cite{eht mag1}-\cite{eht mag3}, the growth
of the hope for the future possible verifications of the other black hole characteristics and observations of new physics effects in the vicinity of them, can be observed. 

From both theoretical and observational points of views the regions of spacetime,
 where the photon orbits are closed, forming timelike hypersurfaces on which bending angle of light is unrestrictedly high, attract much
the attention to. In essence one can suppose that compact objects like black holes, wormholes, neutron stars are surrounded by {\it photon spheres}.

What's more, general relativity and its generalizations 
foresee the existence of such kind of regions. 
The concept of {\it photon sphere} and {\it photon surface} are of a great importance in studies of black hole shadows \cite{vir00,cla01},
 triggered by the recent achievements concerning black hole images, at the center of Milky Way and M87 galaxies \cite{eht1,eht2}.
They are also used in search for the traces of new physics beyond the Standard Model.

In four-dimensional spacetime, {\it photon sphere} special properties,
which recall features of a black hole event horizon, allow us to classify their spacetimes in terms of their asymptotical charges. It constitutes
the alternative for black hole uniqueness theorem \cite{heu96}-\cite{yaz21}. 
Consequently, the generalizations of the uniqueness theorem for $n$-dimensional gravity were also investigated.
In \cite{ced21} the higher-dimensional problem of {\it photon sphere} and uniqueness of higher-dimensional Schwarzschild spacetime was elaborated, while
the {\it photon sphere} uniqueness for electro-vacuum $n$-dimensional spacetime was given in \cite{jah19}. 
On the other hand, in \cite{bud20} the studies of the so-called
{\it trapped photons}, i.e., photons which never pass the event horizon or escape towards spatial infinity, in the spacetime of higher-dimensional Schwarzschild-Tangherlini black hole, have
been elaborated.

The aforementioned concepts of {\it photon sphere} and {\it surface} play the key role in studies Penrose inequalities \cite{shi17}-\cite{yan20}.
It turns out that they are timelike totally umbilic hypersurfaces with the proportionality between their first and second fundamental forms.
Other mathematical and geometrical aspects of these objects were scutinized both for static and stationary axisymmetric spacetimes \cite{gib16}-\cite{kob22}.

On the other hand, the generalization of the {\it photon sphere} concept to the case of massive charged particle.
They describe the case of timelike hypersurfaces to which any wordline
of particles initially touching to them remains in the hypersurface in question \cite{kob22a,bog23}.

As was mentioned above, the studies of {\it photon sphere} properties reveal that it is totally umbilical hypersurface (i.e., its second fundamental form is a pure trace)
with constant mean curvature and surface gravity, being strongly resembled to black hole event horizon. On the other hand,
from black hole theory one knows that the presence of black hole event horizon enables to classify asymptotically flat spacetimes in terms of their asymptotic charges 
(authorizes the uniqueness theorems for various kind of black hole solutions).

Therefore the tantalizing question arises, if the presence of {\it photon sphere} delivers a new tool for the classification of spacetimes with asymptotical charges
or other  physical quantities.

Our paper is concerned with the problem of classification of static asymptotically flat 
spacetimes being the solution of 
Einstein-Maxwell gravity with electric $Q_{(F)}$ and magnetic $Q_{(F)}$ charges, having the line element of the form
\be
ds^2 = - \Bigg( 1 -\frac{2M}{r} + \frac{Q_{(F)}^2 +Q_{(B)}^2}{r^2} \Bigg) dt^2 + \frac{dr^2}{\Big( 1 -\frac{2M}{r} + \frac{Q_{(F)}^2 +Q_{(B)}^2}{r^2} \Big) }
+ r^2 d\Omega^2,
\ee
where $d\Omega^2$ is the metric of the unit sphere, containing
a {\it photon sphere}. The paper is a continuation of our previous work \cite{rog16},
devoted to the uniqueness of {\it photon sphere} for Einstein-Maxwell-dilaton black hole solutions with an arbitrary coupling constant.
In addition the influence of magnetic field on the {\it photon sphere} region is very interesting due to the measurements and observations of black hole magnetic field
by EHT Collaboration and in the context of future planned experiments \cite{eht mag1}-\cite{eht mag3},  \cite{dal18}.


The organization of the paper is as follows. In Sec. II we describe the basic features of Maxwell gauge field in spacetime with the presence
of asymptotically timelike Killing vector field orthogonal to the hypersurface of constant time. Sec. III will be devoted to the basic characteristics of {\it photon sphere}
with electric and magnetic charges. We found the functional dependence among lapse function and aforementioned charges, which will be
of the key importance for revealing that the {\it photon sphere} has scalar constant curvature.  Sec. IV is connected with the basic steps
in proof of the uniqueness theorem, using the conformal positive energy theorem. The alternative way of obtaining the classification (uniqueness) of
non-extremal static asymptotically flat solution in Einstein-Maxwell gravity with electric and magnetic charges will be presented in Sec. V.
The proof is based on method which use the appropriate conformal transformation and positive energy theorem. 
In the last Sec. we conclude our investigations.

\section{Equations of motion with the presence of stationary Killing vector field}
In this section we recall the basic features of electric and magnetic Maxwell fields in the presence of asymptotically timelike Killing vector field.
In what follows we shall consider the ordinary Einstein-Maxwell system given by the action
\be
S_{EM} = \int  d^4x  \sqrt{-g} \Big( R
- F_{\mu \nu} F^{\mu \nu} \Big),
\label{ac}
\ee  
where $g$ is the determinant of the four-dimensional metric tensor, 
$ F_{\mu \nu} = 2 \na_{[\mu} A_{\nu ]}$ stands for the  $U(1)$-gauge field strength.
Variation of the action (\ref{ac}) with respect to metric tensor $g_{\mu\nu}$ and $A_\mu$, reveals the standard form of Einstein-Maxwell equations of motion
\be
\na_{\mu} F^{\mu \nu } = 0, \qquad R_{\mu \nu} = T_{\mu \nu}(F),
\ee
where the energy momentum tensors, defined as $T_{\mu \nu} = - \delta S/ \sqrt{-g} \delta g^{\mu \nu} $, is provided by
\be
T_{\mu \nu}({F}) = 2 F_{\mu \rho} F_{\nu}{}^{\rho} - \frac{1}{2} g_{\mu \nu} F^2,
 \ee
We introduce an asymptotically timelike Killing vector field $k_{\delta}$
and assume that the field strength in the considered theory will be stationary, i.e., 
$\cL_k~F_{\alpha \beta} = 0.$
The exact form of the energy-momentum tensor $T_{\alpha \beta}(F)$ envisages that it also fulfils the stationarity assumption,
$\cL_k~T_{\alpha \beta}(F) = 0$.
The existence of stationary Killing vector field $k_a$ enables one to introduce into consideration the meaning of
the twist vector $\omega_a$, defined as
\be
\omega_a = \frac{1}{2}~\ep_{abcd}~k^b~\na^c~k^d.
\ee
Furthermore, for any Killing vector field one has
$\na_\alpha~\na_\beta \chi_\ga = - R_{\beta \ga \alpha}{}{}^{\delta}~\chi_\delta$, which implies in turn the relation of the form as
 \be
\na_\beta ~\omega_\alpha = \frac{1}{2}~\ep_{\alpha \beta \ga \delta}~k^{\ga}~R^{\delta \chi}~k_{\chi}.
\label{rrq}
\ee
It can be also found that 
$
\na_\alpha~\Big( \frac{\omega^\alpha}{N^4} \Big) = 0,$ 
 where we set $N^2 = - k_\ga~k^\ga$.

The introduction of Killing vector field in question, allows to define
electric and magnetic components for gauge field strengths $F_{\alpha \beta}$ as follows:
\be
E_{\alpha} = - F_{\alpha \beta}~k^\beta, \qquad
B_{\alpha} = \frac{1}{2}~\ep_{\alpha \beta \ga \delta}~k^\beta~F^{\ga \delta},
\ee
and consequently the field strength $F_{\alpha \delta}$ can be rewritten in terms of $E_{\alpha}$ and $B_{\alpha}$, i.e.,
$N^2~F_{\alpha \beta}= - 2~k_{[\alpha} E_{\beta]}+ 
\ep_{\alpha \beta \ga \delta}~k^\ga~B^{\delta}.$
On the other hand, the equations of motion
for magnetic and electric parts gauge field strength are provided by
\ben \label{prop1}
\na_{\alpha}\bigg( {E^{\alpha} \over N^2} \bigg) &=& 2~{B^{\ga} \over N^4}~\omega_\ga,\\ \label{prop2}
\na_{\alpha}\bigg( {B^{\alpha} \over N^2} \bigg) &=& - 2~{E^{\ga} \over N^4}~\omega_\ga,
\een
while the field invariance conditions
$\cL_k~F_{\alpha \beta}= 0$, as well as, the
relations $\na_{[ \ga} F_{\alpha \beta ]} = 0$, establish the generalized Maxwell source-free equations in the form
\be
\na_{[\alpha}E_{\beta ]} = 0,\qquad
\na_{[\alpha}B_{\beta ]} = 0.
\ee
Our considerations we shall consider the simply connected spacetime. It permits us to implement the electric and magnetic potentials in the forms as
\be
E_{\alpha} = \na_{\alpha} \psi_{F}, \qquad
B_{\alpha} = \na_{\alpha} \psi_B.
\ee
Having in mind
the relations (\ref{rrq}) and the explicit form
of the Ricci tensor, one finds the {Poynting flux} in Einstein-Maxwell gravity with electric and magnetic charges. It yields
\be
\na_{[\alpha} \omega_{\beta ]} = 4~ E_{[ \alpha} B_{\beta ]}.
\label{oom}
\ee 
In what follows we shall pay attention to static spacetime, i.e., one supposes that there exists a smooth Riemannian manifold and a smooth lapse function $N: M^3 \rightarrow R^{+}$, such that 
$M^4 = R \times M^3$. The assumptions provide that the line element in the studied spacetime can be written in the form as
\be 
ds^2 =  g_{\mu \nu} dx^\mu dx^\nu = - N^2 dt^2 + g_{ab} dx^a dx^b,
\label{metric}
\ee
where $N$ and $g_{ab}$ are time independent, as they are determined on the hypersurface of constant time.

The spacetime under consideration is asymptotically flat containing a data set $(\Sigma_{end},~g_{ij},~K_{ij})$ with gauge fields $A_\mu$ such that 
$\Sigma_{end}$ constitutes a manifold diffeomorphic to $R^{(3)}$ minus a 
closed 
unit ball at the origin of $R^{(3)}$. Besides, it is subject to the following
asymptotic behaviors of $g_{ij},~F_{\mu \nu}$:
\ben
\mid g_{ij} &-& \delta_{ij} \mid + r~ \mid \p_a g_{ij} \mid + \dots + r^k ~\mid \p_{a_1 \dots a_k} g_{ij} \mid \\ \nonumber
&+& r \mid K_{ij} \mid
+ \dots + r^k~ \mid \p_{a_1 \dots a_k} K_{ij} \mid \le \cO \big( \frac{1}{r} \Big), \\
F_{ \alpha \beta} &+& r ~\mid \p_a F_{\alpha \beta} \mid + \dots + r^k \mid \p_{a_1 \dots a_k} F_{\alpha \beta}\mid  \le \cO \Big(\frac{1}{r^2} \Big).
\een

The Einstein-Maxwell equations are provided by the following:
\ben \label{1emd}
{}^{(g)}\na_i {}^{(g)}\na^i N &=& \frac{1}{N} \Big( {}^{(g)}\na_i \psi_{F} {}^{(g)}\na^i \psi_{F} + {}^{(g)}\na_i \psi_{B} {}^{(g)}\na^i \psi_{B} \Big),\\ \label{2emd}
N~{}^{(g)}\na_i {}^{(g)}\na^i \psi_{F} &=& {}^{(g)}\na_i N~{}^{(g)}\na^i \psi_{F},\\
N~{}^{(g)}\na_i {}^{(g)}\na^i \psi_{B} &=& {}^{(g)}\na_i N~{}^{(g)}\na^i \psi_{B},\\
{}^{(g)} R &=& \frac{1}{N^2} \Big( 
{}^{(g)}\na_i \psi_{F}~{}^{(g)}\na^i \psi_{F} + {}^{(g)}\na_i \psi_{B}~{}^{(g)}\na^i \psi_{B} \Big),\\ \nonumber
{}^{(g)} R_{ij} &=& \frac{1}{N} {}^{(g)}\na_i {}^{(g)}\na_j N +
\frac{1}{N^2} \Big[
g_{ij} \Big( {}^{(g)}\na_k \psi_{F} {}^{(g)}\na^k \psi_{F} + {}^{(g)}\na_k \psi_{B} {}^{(g)}\na^k \psi_{B} \Big) \\ \label{3emd}
&-& 2 ~\Big( {}^{(g)}\na_i \psi_{F} {}^{(g)}\na_j \psi_{F} + {}^{(g)}\na_i \psi_{B} {}^{(g)}\na_j \psi_{B} \Big) \Big],
\een
where ${}^{(g)}\na_i $ is the covariant derivative with respect to metric tensor $g_{ij}$.
${}^{(g)} R_{ij}$ denotes the three-dimensional Ricci tensor, while ${}^{(g)} R$ stands for the Ricci scalar curvature.

\section{Geometry of photon sphere in static asymptotically flat spacetime with electric and magnetic potentials}
This section will be devoted to the description {\it photon sphere} with one single component \cite{yaz15b}. 
Namely,
a {\it photon surface} is an embedded timelike hypersurface for which any null geodesics initially tangent to it, remains tangent during 
the passage of time of its existence. On the other hand, by {\it photon sphere} one defines a photon surface with constant lapse function $N$, and the additional 
conditions imposed on the electric and magnetic charges emerging in the studied theory. 

We suppose that the lapse function regularly foliates the manifold outside the {\it photon sphere}. Thus it effects that all level
sets with $N= const$ are topological spheres. It yields that outside {\it photon sphere} one has $1/\rho^2 = {}^{(g)}\na_i N {}^{(g)}\na^i N \neq 0$.

Further, we define electric and magnetic static system as a time slice of the static spacetime 
$(R \times M^3, -N^2 dt^2 + g_{ij}dx^i dx^j)$. Then, we define the notion of the {\it photon surface}. Namely, let
$(M^3,~g_{ij},~N,~\psi_{F},~\psi_{B})$ be a Maxwell electric magnetic system bounded with
spacetime defined above, with the metric (\ref{metric}). By the {\it photon sphere} 
we shall understand a timelike hypersurface embedded
$(P^3,~h_{ij}) \hookrightarrow (R \times M^3, -N^2 dt^2 + g_{ij}dx^i dx^j)$. If the embedding is umbilic and the gradient of the lapse function, the electric 
 one-form is normal to
$P^3$.

\subsection{Mean curvature of photon sphere}
It has been revealed in \cite{cla01}  that
the second fundamental form of $P^3$ may be written as 
$
K_{ij} = \frac{1}{3}\Theta h_{ij},
$
 where $\Theta$ stands for the expansion of the unit normal to the {\it photon sphere}. Moreover, the condition for a timelike hypersurface to be a {\it photon sphere} 
 is its  total umbilicity (its second fundamental form is a pure trace).

To commence with, we use the Codazzi equations to analyze the properties of the {\it photon sphere} in question.
 We denote by $n_a$ unit normal to the {\it photon
sphere}, $Y_\beta$ will represent the element of tangent space $TP^3$ 
and using these quantities one obtains that for all vectors vectors $Y_c$ belonging to $TP^3$, the following relation is satisfied:
\be
\frac{1}{3} \Theta_{,b}~(1 -3)~Y^b = {}^{(g)}R_{cd}~n^c~Y^d,
\label{the}
\ee
where the right-hand side of equation (\ref{the}) is given by
\ben
{}^{(g)}R_{cd}~n^c~Y^d &=& 2 \frac{1}{N^4}~k_a~Y^a~k_b~n^b~\Big( E_m E^{m} + B_m B^{m} \Big) \\ \nonumber
&-&
2 \frac{1}{N^2} \Big( E_a E_b + B_a B_b \Big)
Y^a~ n^b + \frac{n_k Y^k}{N^2} \Big( E_m E^{m} + B_m B^{m} \Big).
\een
Having in mind the fact that electric fields $E_a,~B_a$  ($E_a$ is normal to $P^3$, by its definition
and the results of subsection C
show that this is the case for $B_a$ in static spacetime)
are normal to the $P^3$ and $k_\alpha Y^\alpha = 0,~ k_\beta n^\beta = 0$,
one concludes that ${}^{(g)}R_{cd} n^c Y^d $ is equal to zero, and we arrive at
\be
0 = (1 -3)~Y^\zeta~\frac{ \Theta_{,\zeta}}{3}.
\ee
Thus for an arbitrary vector $Y^\beta$, the mean curvature of the considered {\it photon sphere} is constant.\\
It can be also shown  \cite{yaz15b} that $\cL_X (n^j {}^{(g)}\na_j N) =0$, where $X$ is an arbitrary tangent vector to $\Sigma^2$, envisaging that
$n^j {}^{(g)}\na_j N$ is constant on $\Sigma^2$.

\subsection{Scalar curvature of electric-magnetic photon sphere}
The scalar curvature of the {\it photon sphere} in question will be found by means of the
contracted Gauss equation. It implies
\be
{}^{(g)}R - 2~{}^{(g)}R_{ij}n^in^j  = {}^{(p)}R - \frac{2}{3} \Theta^2,
\ee
where in our case ${}^{(g)}R_{ij}n^in^j $ yields
\be
{}^{(g)}R_{ij}n^in^j  = - \frac{1}{N^2} \Big( E_a E_b + B_a B_b \Big) n^a n^b.
\ee
As a result, one achieves the relation for the scalar curvature of the {\it photon sphere} 
\be
{}^{(p)}R = \frac{2}{3} \Theta^2 + 2~\frac{1}{N^2} \Big( E_a E_b + B_a B_b \Big) n^a n^b.
\ee
In order to show that {\it photon sphere} has constant scalar curvature one needs to prove that $E_a n^a$ nad $B_k n^k$ are constant on $P^3$.
Above we mentioned that $n^j {}^{(g)}\na_j N$ is constant on $P^3$ (see for the proof \cite{yaz15b}), in the next subsection one envisages that
electric and magnetic potentials are function of $N$, see relation (\ref{funcn}). It leads to the conclusion that $E_a n^a$ nad $B_k n^k$ are constant on $P^3$, implying that
$P^3$ has the constant scalar curvature.

\subsection{Functional dependence - lapse function electric and magnetic potentials}
\label{func}
In static spacetime with Killing vector field $k_\mu$, one has that the twist vector $\omega_\alpha$  (\ref{oom}) is equal to zero. It implies that
proportionality between magnetic and electric fields \cite{heu96}. Due to the fact that
electric one-form is spacelike ($k_\mu$ is timelike), every one-form parallel and orthogonal to it vanishes (\ref{prop1})-(\ref{prop2}). Moreover having in mind the asymptotic conditions
$\psi_F \rightarrow 0$
and $\psi_B \rightarrow 0$, when $r \rightarrow \infty$, we get that
\be
\psi_B = \mu~\psi_F,
\label{elmag}
\ee
where $\mu$ is constant.

As in \cite{isr67}, one can introduce coordinates on $N=const,~t=const$ manifold provided by
\be
g_{ab} dx^a dx^b = {}^{(2)}g_{ab} dy^a dy^b + \rho^2 dN^2.
\ee
Having in mind equations of motion for electric and magnetic potentials and the relation (\ref{elmag}), we obtain
\be
\frac{1}{\sqrt{{}^{(2)}g}} \Big[ \sqrt{{}^{(2)}g} \frac{\phi_F}{N} \Big] = - \frac{\Big( \rho~\psi_F^{;a} \Big)_{;a}}{N},\\
\label{in1}
\ee
where we have denoted
\be
\frac{\p \psi_F}{\p N} = \rho~\phi_F, 
\ee
and the gravitational relation of the form
\be
\frac{1}{\rho^2} \frac{\p \rho}{\p N} = K + \frac{2 \rho (1+\mu^2)}{N}~\Big( \phi_F^2  
+ \psi_{F;a} \psi_F^{;a} 
\Big),
\label{in2}
\ee
where $K = K_m{}^m$ is the the extrinsic scalar curvature of $N=const$ spacetime.
Based on equations of motion (\ref{in1})-(\ref{in2}) for the theory in question, one can arrive at the integral identity given by
\ben \label{iquality}
\frac{1}{\sqrt{{}^{(2)}g}}
 \frac{\p}{\p N}  \Big[
 \sqrt{{}^{(2)}g}
  \Big( \frac{1}{N}
F(N, \tpsi ) \tphi &+& \frac{G(N, \tpsi )}{\rho} \Big) \Big] \\ \nonumber
= A~\rho~\Big( \tphi^2 + \tpsi_{; a} \tpsi^{;a} \Big) + C~\tpsi  &+& \frac{1}{\rho} \frac{\p G}{\p N}
- \frac{1}{N} \Big( F~\rho~\tpsi^{;a} \Big)_{;a},
\een
for differentiable 
arbitrary (for the time being) functions 
$F$, $G$ and the new potential $\tpsi = \sqrt{1 + \mu^2}~ \psi_F$ (where we have used the dependence of electric and magnetic potentials in the static spacetime),
for the same reason we get that $\tphi = \sqrt{1 + \mu^2}~ \phi_F.$
The function $A$ and $B$ are provided by
\ben
A &=& \frac{1}{N} \Big( G + \frac{\p F}{\p \tpsi} \Big) ,\\
B &=& \frac{1}{N} \frac{\p F}{\p N} + \frac{\p G}{\p \tpsi}.
\een
In order to achieve to the integral conservation laws 
from (\ref{iquality})
, we have to restrict our consideration to the case when
$ A = B = \frac{\p G}{\p N} = 0$. The general solutions of the above 
over-determined linear system of differential equation for $F$ and $G$ constitute a linear combination of the
following particular solutions:
\be
F =1,~ G=0, \qquad F = 2 \tpsi,~ G=1, \qquad F = 2\tpsi^2 - N^2,~G = 2 \tpsi.
\ee 
One can integrate the relation (\ref{iquality}), with respect to the all aforementioned values of functions $F$ and $G$, having in mind that the integral 
of two-dimensional divergence over a closed $N=const$ space disappears.
The two boundary surfaces $\Sigma_0$ and $\Sigma_\infty$ were taken into account with the appropriate asymptotic conditions imposed on fields
and characteristic features of them. Namely, for approaching to $\Sigma_\infty$ one has that $r \psi_F  \rightarrow  Q_{(F)},~ r^2 \phi_F \rightarrow -Q_{(F)},
~\frac{\rho}{r^2} \rightarrow \frac{1}{M}$. For $\Sigma_0$ we have that $\phi_F = \cO(N),~\psi_{F;a} = \cO(N)$. On $\Sigma_0$ $\psi_F$ and $1/\rho$ are constant 
\cite{isr67}.

Finally one arrives at the following:
\ben
&{}& \int_{\Sigma_0} dS \Big( \frac{\phi_F}{N} \Big) = - Q_{(F)},\\
2 \Big( 1 &+& \mu^2 \Big) \psi_{(0) F} \int_{\Sigma_0} dS \Big( \frac{\phi_F}{N} \Big) + \frac{S_0}{\rho_0} = M,\\
2 \Big( 1 &+& \mu^2 \Big) \psi_{(0) F}^2 \int_{\Sigma_0} dS \Big( \frac{\phi_F}{N} \Big) + 2 \frac{S_0}{\rho_0} \psi_{(0) F} = Q_{(F)},
\een
where $S_0$ is the area of two-space $\Sigma_0$.\\
It can be seen that the addition of magnetic charge does not change the basic features of {\it photon sphere} (as obtained in Maxwell case in \cite{yaz15b}).
Qualitative features like the constancy of its mean curvature and scalar curvature are the same, however quantitive ones are different.
Namely, they are valid for the modified
potential $\tpsi = \sqrt{1 + \mu^2}~ \psi_F$, on which magnetic potential imprints its influence.

All the above reveal that one arrives at the following functional dependence among $N_0$ lapse function on $\Sigma_0$, ~$\psi_{(0) F}$ electric potential at
$\Sigma_0$ and the constant $\mu$ bounded magnetic and electric potentials:
\be
 2 \Big( 1 + \mu^2 \Big) \psi_{(0) F}^2 + 2 \psi_{(0) F} \frac{M}{Q_{(F)}} - 1 = N_0^2,
\label{funcn}
 \ee
 as was mentioned above $\psi_{(0) F}$ and $N_0$ are constant on the considered hypersurface and $\psi_F \rightarrow 0$, as $r \rightarrow \infty$. 

The equations (\ref{funcn}) is valid not only on the surface in question but also in all its exterior region. Namely, let us compose the divergence identity based on the 
above equations
\be
\frac{1}{2} {}^{(g)}\na^m \Bigg[\Big(-N^2 + 2 (1+\mu^2) \psi_F^2 + \frac{2 \psi_F M}{Q_{(F)}} -1\Big) ~\theta_m \Bigg] = N~\theta_m \theta^m,
\label{gauss}
\ee
where $\theta^m$ yields
\be
\theta^m = - {}^{(g)} \na^m N + \frac{1}{N} \Big( 2(1+\mu^2) \psi_F~{}^{(g)} \na^m \psi_F + \frac{M}{Q_{(F)}} {}^{(g)} \na^m \psi_F \Big).
\ee
In the next step one applies the Gauss theorem to the relation (\ref{gauss}), taking into account the asymptotic behaviors of $N, ~\psi_F$, and the
fact that $N>0$ in the exterior region of {\it photon sphere}, one can draw a conclusion that $\theta^m = 0$. Fixing in this relation the integration constant as equal to 1,
we arrive at the equation expressing a functional dependence among electric/magnetic potentials and  $N$.

It proves the constancy of $E^a n_a$ and $B_c n^c$ on $P^3$, implying that ${}^{(g)} R$
is a constant scalar curvature.

\subsection{Auxiliary formulae}
Some additional formulae envisaging the influence of magnetic charge on the {\it photon sphere} can be obtained
 the equation of motion (\ref{1emd}), for the isometric embedding $(\Sigma^2,~\sigma_{ij}) \hookrightarrow (M^3,~g_{ij})$. 
 Namely, if one considers
the contracted Gauss relation, it yields
\be
N~{}^{(\sigma)} R =  \frac{2}{N} \Big( E_a E_b + B_a B_b \Big) n^a n^b
+ 2 H~n^k~{}^{(g)}\na_k N + \frac{H^2}{2} N,
\label{kkk}
\ee
where we have denoted $H = \frac{2}{3} \Theta$.
Integration  of (\ref{kkk}) over the hypersurface $\Sigma$ results in
\ben
\int_\Sigma d\Sigma~N~{}^{(\sigma)} R &=& \int_\Sigma ~d\Sigma~\frac{2}{N} \Big( E_a E_b + B_a B_b \Big) n^a n^b\\ \nonumber
&+& 2 \int_\Sigma d\Sigma~H~n^k~{}^{(g)}\na_k N + \int_\Sigma d\Sigma~\frac{H^2}{2} N.
\een
Let us examine the area of the hypersurface $\Sigma$ denoted by $A_\Sigma$, and apply the Gauss-Bonnet theorem.
Consequently we arrive at
\be
N_0 = \frac{1}{4\pi~N_0} \Big( E_a E_b + B_a B_b \Big) n^a n^b
A_\Sigma + H~M_{Phs} + \frac{1}{16 \pi} H^2~A_\Sigma~N_0,
\label{no}
\ee
where the mass of the {\it photon sphere} implies
\be
M_{phs}  = \frac{1}{4 \pi} n^k~{}^{(g)}\na_k N ~A_\Sigma.
\ee

Next, we take into account the contracted 
Gauss equation ${}^{(\sigma)}R = {}^{(p)}R - 2~{}^{(p)}R_{ij} \eta^i \eta^j$,
for $(\Sigma^2,~\sigma_{ij}) \hookrightarrow (P^3,~h_{ij})$ isometric embedding, with a unit normal $\eta_i$.

The same procedure as above,
leads us to the equation provided by
\be
1 = \frac{3}{16 \pi} H^2~A_\Sigma + \frac{1}{4 \pi~N_0^2}  \Big( E_a E_b + B_a B_b \Big) n^a n^b
 ~ A_\Sigma,
\label{nnn}
\ee
which can be rewritten using the definition of the electric and magnetic charges
\be
Q_{(F)}= - \frac{A_\Sigma~E_k n^k}{4 \pi~N_0}, \qquad Q_{(B)} = - \frac{A_\Sigma~B_k n^k}{4 \pi~N_0},
\ee
and equations (\ref{no})  and (\ref{nnn}), in the form as
\be
1 = \frac{4 \pi \Big( Q_{(F)}^2 + Q_{(B)}^2\Big)}{A_\Sigma} + \frac{3}{2}~\frac{H}{N_0}~M_{phs}.
\ee
On the other hand, 
the relation among $n^a,~N_0,~H$ yields
\be
2 ~n^a {}^{(g)} \na_a N = H~N_0.
\label{nn}
\ee
Then, using (\ref{nn}) we get the expression envisaging how magnetic charge influences $A_\Sigma$.
\be
\frac{A_\Sigma}{4 \pi} = \Big( Q_{(F)}^2 + Q_{(B)}^2\Big) + 3 \frac{M_{phs}^2}{N_0^2}
= \Big(1 + \mu^2\Big) Q_{(F)}^2  + 3 \frac{M_{phs}^2}{N_0^2}.
\ee

\section{Uniqueness of photon sphere with electric and magnetic charges}

The uniqueness proof of the {\it photon sphere} in Einstein-Maxwell {\it dark photon} gravity will be conducted in  several steps, having in mind the attitude
presented in Refs. \cite{ced15a}-\cite{rog16}. 

In this section the {\it photon sphere} emerges as the inner boundary of the studied spacetime \cite{ced15}. Namely one has that
\be
(P^3,~h_{ij}) 
= \cup_{i=1}^I~(R \times \Sigma_i^2,~-N_i^2 dt^2 + \sigma^{(i)}_{ij}dx^i dx^j),
\ee
where $P_i^3$  denotes each connected component of $P^3$. 

To begin with, we define
 the electric-magnetic Einstein-Maxwell {\it dark photon} system as \\
$(M^3,~g_{ij}, ~N, ~\psi_{F},~\psi_{B})$, being asymptotic to static spherically symmetric solution in the considered theory, and possessing
Killing horizon boundary. It can be done by glueing spatial pieces of the aforementioned solution, having the adequate mass and Maxwell electric and magnetic
charges. Namely, each of the {\it photon sphere} $\Sigma_i^2$ will be glued at neck piece of Einstein-Maxwell manifold with mass greater than zero and having charges $Q_i^{({F})},~Q_i^{({B})}$.

The manifold $M^3$ will be smooth, the metric tensor, lapse function, and the potentials $\psi_{F},~\psi_{B}$ 
will be smooth away from the glueing surface. The considered manifold will be characterized by non-negative scalar curvature (away from the glueing surface)
 and will be geodesically complete.

Next the adequate conformal transformations will be applied and we use the conformal positive energy theorem 
to show that the {\it photon sphere}
is isometric to Einstein-Maxwell spacetime
characterized by the Arnowitt-Deser-Misner (ADM) mass, electric and magnetic charges. 
The non-degenerate case of Einstein-Maxwell static, spherically symmetric system will be considered.

\subsection{Conformal positive theorem and uniqueness of electric-magnetic photon sphere}
In our attitude to the problem, the conformal positive energy theorem, derived in Refs. \cite{sim99, gib02a}, will account for the key role in the
uniqueness proof.
In order to implement it to our considerations one ought to fulfil its assumptions. Namely, 
 we have to consider two asymptotically flat 
Riemannian $(n-1)$-dimensional manifolds, $(\Sigma^{(\Phi)},~ {}^{(\Phi)}g_{ij})$ and $(\Sigma^{(\Psi)},~ {}^{(\Psi)}g_{ij})$ for
which metric tensors are bounded with the conformal transformation given by
\be
{}^{(\Psi)}g_{ij} = \Omega^2~{}^{(\Phi)}g_{ij}, 
\ee
where $\Omega$ stands for a conformal factor. 
On the other hand, the masses of the above manifolds satisfy
the relation of the form
${}^{(\Phi)}m + \beta~{}^{(\Psi)}m \geq 0$, under the auxiliary requirement putting on the Ricci curvature scalar tensor
${}^{(\Phi)} R + \beta~\Omega^2~{}^{(\Psi)} R \geq 0$, 
where ${}^{(\Phi)} R $ and ${}^{(\Psi)} R$ are the Ricci scalars with respect to the adequate metric tensors, defined on the two manifolds, while
$\beta$ is a positive constant. The inequalities are satisfied if
$(n-1)$-dimensional manifolds are flat \cite{sim99}. 

The conformal positive energy theorem was widely applied in proving the uniqueness of four and higher-dimensional black objects \cite{mar02}-\cite{rog22} and
wormhole solutions \cite{wormholes18}.

In our considerations we implement the conformal transformation of the form as follows:
\be
\tilde g_{ij} = N^{2} g_{ij},
\ee
leading to the conformally rescaled Ricci tensor given by
\be
\tR_{ij}(\tg) = \frac{2}{N^2} {}^{(g)}\na_i N~{}^{(g)}\na_j N - \frac{2}{N^2}\Big( {}^{(g)}\na_i \psi_{F} {}^{(g)}\na_j \psi_{F} + {}^{(g)}\na_i \psi_{B} {}^{(g)}\na_j \psi_{B} \Big) 
\ee

Next, 
we define the quantities provided by the relations, for electric potential $\psi_F$
\ben \label{p1}
\Phi_{1}&=& \frac{1}{2} \Big( N + \frac{1}{N} - \frac{2}{N}~ \psi_{F}^2 \Big),\\
\Phi_0 &=& \frac{\sqrt{2}}{N} ~\psi_{F},\\
\Phi_{-1} &=& \frac{1}{2} \Big( N - \frac{1}{N} - \frac{2}{N}~ \psi_{F}^2 \Big),
\een
and the quantities including $\psi_B$ potential are given by
\ben
\Psi_{1}&=& \frac{1}{2} \Big( N + \frac{1}{N} - \frac{2}{N} \psi_{B}^2 \Big),\\
\Psi_0 &=& \frac{\sqrt{2}}{N} \psi_{B},\\ \label{ps3}
\Psi_{-1} &=& \frac{1}{2} \Big( N - \frac{1}{N} - \frac{2}{N} \psi_{B}^2 \Big).
\een
It can be observed that the auxiliary constraint relation can be found when one defines the metric tensor $\eta_{AB} = diag(1, -1, -1)$. They are provided by
that
\be
\Phi_A \Phi^A = \Psi_A \Psi^A = -1.
\label{ffpp}
\ee
where we set $A = - 1, 0, 1$.
Consequently, the other symmetric tensors can be constructed, for the potential $\Phi_A$
\be 
\tG_{ij} = \tna_{i} \Phi_{-1} \tna_{j} \Phi_{-1} - \tna_{i} \Phi_{0} \tna_{j} \Phi_{0} - \tna_{i} \Phi_{1} \tna_{j} \Phi_{1},
\label{g1}
\ee
and similarly for the potential $\Psi_{A}$
\be
\tH_{ij} = \tna_{i} \Psi_{-1} \tna_{j} \Psi_{-1} - \tna_i \Psi_{0} \tna_j \Psi_0- \tna_{i} \Psi_{1} \tna_{j} \Psi_{1},
\label{h1}
\ee
where $\tna_{i}$ denotes for the covariant derivative with respect to the conformally rescaled metric $\tg_{ij}$.

Due to the relations (\ref{ffpp}) one arrives at
\be
\tna^{2}\Phi_{A} = \tG_{i}{}{}^{i} \Phi_{A}, \qquad
\tna^{2} \Psi_{A} = \tH_{i}{}{}^{i} \Psi_{A}.
\label{ppff}
\ee
Moreover, the Ricci curvature tensor $\tR_{ij}$ connected with conformally rescaled metric $\tg_{ij}$
may be rewritten in terms of $\tG_{ij}$ and $\tH_{ij}$, i.e.,
\be
\tR_{ij} =  \tG_{ij} + \tH_{ij}.
\label{rr}
\ee
The relations (\ref{ppff}) and (\ref{rr}) can be derived by varying the Lagrangian density \cite{mar02, hoe76, sim92}
\be
\cL = \sqrt{-\tg} \Big( \tG_{i}{}^{i} + \tH_{i}{}^{i} + \frac{ \tna^i  \Phi_{A} \tna_i  \Phi^{A}}{ \Phi_{A} \Phi^{A}} + \frac{ \tna^i  \Psi_{A} \tna_i  \Psi^{A}}{ \Psi_{A} \Psi^{A}} \Big),
\ee
with respect to $\tg_{ij},~ \Phi_{A}, ~ \Psi_{A}$, and taking into account the constraint relations (\ref{ffpp}).

The conformal positive energy theorem account for the main point in the uniqueness theorem.
Thus, we suppose that one has two asymptotically flat Riemannian three-dimensional manifolds
$(\Sigma^{\Phi},~ {}^{(\Phi)}g_{ij})$ and $(\Sigma^{\Psi},~ {}^{(\Psi)}g_{ij})$.
 The conformal transformation between two manifolds will be of the form as
${}^{(\Psi)}g_{ij} = \Omega^2~{}^{(\Phi)}g_{ij}$. It provides that the corresponding masses satisfy
the relation ${}^{\Phi}m + \beta~{}^{\Psi}m \geq 0$ if ${}^{(\Phi)} R + \beta~\Omega^2~{}^{(\Psi)} R \geq 0$, for some
positive constant $\beta$. The inequalities in question ensure that the
three-dimensional Riemannian manifolds are flat.

In order to satisfy the requirement of
the conformal positive energy theorem, we introduce into considerations
conformal transformations provided by
\be
{}^{(\Phi)}g_{ij}^{\pm} = {}^{(\Phi)}\omega_{\pm}^{2}~ \tg_{ij},
\qquad
{}^{(\Psi)}g_{ij}^{\pm} = {}^{(\Psi)}\omega_{\pm}^{2}~ \tg_{ij},
\ee
where the conformal factors imply
\be
{}^{(\Phi)}\omega_{\pm} = {\Phi_{1} \pm 1 \over 2}, \qquad
{}^{(\Psi)}\omega_{\pm} = {\Psi_{1} \pm 1 \over 2}.
\label{pf}
\ee
Then, the standard procedure of pasting of three-dimensional manifolds $(\Sigma_{\pm}^{\Phi},~ {}^{(\Phi)}g_{ij}^{\pm})$ 
and 
$(\Sigma_{\pm}^{\Psi},~ {}^{(\Psi)}g_{ij}^{\pm})$ across their shared minimal boundary, can be put into application.
As a result we obtain
four manifolds $(\Sigma_{+}^{\Phi},~ {}^{(\Phi)}g_{ij}^{+})$,
$(\Sigma_{-}^{\Phi},~ {}^{(\Phi)}g_{ij}^{-})$,~ $(\Sigma_{+}^{\Psi},~ {}^{(\Psi)}g_{ij}^{+})$,~$(\Sigma_{-}^{\Psi},~ {}^{(\Psi)}g_{ij}^{+})$,
which will be pasted
across shared minimal boundaries ${B}^\Psi$ and ${B}^\Phi$.
Thus, complete regular hypersurfaces
$\Sigma^{\Phi} = \Sigma_{+}^{\Phi} \cup \Sigma_{-}^{\Phi}$ and $\Sigma^{\Psi} = \Sigma_{+}^{\Psi} \cup \Sigma_{-}^{\Psi} $, can be constructed.

It can be checked that the total gravitational mass ${}^{\Phi} m$ on hypersurface $\Sigma^{\Phi}$ and ${}^{\Psi}m$
 on $\Sigma^{\Psi}$ vanish, i.e., it can shown that the metric tensors connected with the adequate hypersurface are proportional to Kronecker delta \cite{gib02a}.

 On this account, it is customary to define 
another conformal transformation provided by the relation
\be
{\hat g}^{\pm}_{ij} = \bigg[ \bigg( {}^{(\Phi)}\omega_{\pm} \bigg)^2
 \bigg( {}^{(\Psi)}\omega_{\pm} \bigg)^{2} \bigg]^{1 \over 2}\tg_{ij},
\ee
leading to the following form of
the Ricci curvature tensor on the defined space:
\ben \label{ric}
\hat R_\pm &=& \bigg[ {}^{(\Phi)}\omega_{\pm}^2~ {}^{(\Psi)}\omega_{\pm}^{2 } \bigg]
^{-{1 \over 2}}
\bigg( {}^{(\Phi)}\omega_{\pm}^{2} {}^{(\Phi)}R_\pm +
{}^{(\Psi)}\omega_{\pm}^{2} {}^{(\Psi)}R_\pm \bigg) \\ \nonumber
&+& 
\bigg( \hat \na _{i} \ln {}^{(\Phi)}\omega_{\pm} - {\hat \na} _{i} \ln {}^{(\Psi)}\omega_{\pm} \bigg)  
\bigg( \hat \na ^{i} \ln {}^{(\Phi)}\omega_{\pm} - {\hat \na}^{i} \ln {}^{(\Psi)}\omega_{\pm} \bigg).  
\een
The direct calculations revealed that the first term on the right-hand side of the above equation can be rewritten as
\ben \label{pos1}
{}^{(\Phi)}\omega_{\pm}^{2}~ {}^{(\Phi)}R_\pm + {}^{(\Psi)}\omega_{\pm}^{2}~ {}^{(\Psi)}R_\pm &=& 
2~\mid {\Phi_{0} \tna_{i} \Phi_{-1}
- \Phi_{-1} \tna_{i} \Phi_{0} \over
\Phi_{1} \pm 1 } \mid^2 \\ \nonumber
&+& 2~\mid { \Psi_{0} \tna_{i} \Psi_{-1}
- \Psi_{-1} \tna_{i} \Psi_{0} \over
\Psi_{1} \pm 1} \mid^2.
\een
It leads to the conclusion, that in terms of the equations (\ref{ric}) and (\ref{pos1}),  the Ricci scalar $\hat R_\pm $ is greater or equal to zero.

Consequently, on the account of the conformal positive energy theorem, it is revealed that the manifolds 
$(\Sigma^{\Phi},~ {}^{\Phi}g_{ij})$, $(\Sigma^{\Psi},~ {}^{\Psi}g_{ij})$ and
$(\hat \Sigma,~ {\hat g}_{ij})$ are flat, which in turn enables us to claim that 
${}^{(\Phi)}\omega = const.~{}^{(\Psi)}\omega$,~ $\Phi_{0} = const~ \Phi_{-1}$ and $\Psi_{0} = const~ \Psi_{-1}$.

Just we can concluded that the manifold $(\Sigma,~ g_{ij})$ is conformally flat. Moreover its metric tensor $\hat g_{ij}$ can be rearranged in conformally flat form
$\hat g_{ij} = {\cal U}^{4}~ {}^{(\Phi)}g_{ij},$
where the conformal factor is given by ${\cal U} = ({}^{\Phi}\omega_{\pm} N)^{-1/2}$. 

Having in mind the relation (72), calculating Ricci scalar $\hat R$, we obtain ${}^{(\Phi)}R$ plus term proportional to $\nabla^2 {\cal U}$ \cite{gib02a}. 
Because $\hat R = {}^{(\Phi)}R = 0$, thus
$\cal U$ is harmonic function  on the three-dimensional Euclidean manifold
$
\na_{i}\na^{i}{\cal U} = 0,
$
where $\na$ is the covariant derivative on a flat manifold. 

One can define a local coordinate for the base space in the form
\be
{}^{(\Phi)}g_{ij} dx^{i}dx^{j} = \tilde \rho^{2} d{\cal U}^2 + {\tilde h}_{AB}dx^{A}dx^{B}.
\ee
The {\it photon sphere} will be located at some constant value of ${\cal U}$ and
the radius of the {\it  photon sphere} can be given at the fixed value of
$\rho$-coordinate \cite{ced15a}.
All these enable that on the hypersurface $\Sigma$ the metric tensor can be given in the form of 
$$\hat g_{ij}dx^{i}dx^{j} = \rho^2 dN^2 + h_{AB}dx^{A}dx^{B},$$
and a connected component of the {\it photon surface} can be identified at fixed value of $\rho$-coordinate.

Suppose that
${\cal U}_{1}$ and ${\cal U}_{2}$ comprise
two solutions of the boundary value problem of the Einstein-Maxwell system with electric and magnetic charges.
Having in mind the Green identity, integrating over the volume element, one gets
\be
\bigg( \int_{r \rightarrow \infty} - \int_{\cal H} \bigg) 
\bigg( {\cal U}_{1} - {\cal U}_{2} \bigg) {\p \over \p r}
\bigg( {\cal U}_{1} - {\cal U}_{2} \bigg) dS = \int_{\Omega}
\mid \na \bigg( {\cal U}_{1} - {\cal U}_{2} \bigg) \mid^{2} d\Omega.
\label{green}
\ee
The surface integrals on the left-hand side of the equation (\ref{green})
vanish because of the imposed boundary conditions and provides that the volume  integral have to be identically equal to zero.
It all leads to the conclusion that the considered two solutions of the Laplace equation with the Dirichlet boundary conditions are identical.

\subsection{Positive mass theorem and uniqueness}
For the completeness of the presented results, we propose the alternative way of conducting the uniqueness proof of electric and magnetic charged photon sphere, 
based on the another conformal transformation and use of the positive energy theorem \cite{posen}-\cite{mas92}. Namely consider the conformal transformation
on $(\Sigma,~\Omega^2 g_{ij})$, then one pastes two copies of $\Sigma_\pm$ along the boundary, and takes into account the conformal transformations
on each copy of $\Sigma$, i.e., $\Omega_\pm^2 g_{ij}$. The conformal factors yield \cite{heu96,heu94}
\be
\Omega_{\pm} = \frac{1}{4} \Big[ \Big(1 \pm N \Big)^2 - Z Z^* \Big].
\ee
Ricci curvature for the metric $\Omega^2 g_{ij}$, where for the brevity we denote $\Omega = \Omega_\pm$, has the form as follows:
\ben \label{ricci}
\frac{1}{2} \Omega^4 N^2 ~R(\Omega^2 g_{ij}) &=& \mid \Big( \Omega - N \frac{\p \Omega}{\p N} \Big) {}^{(g)}\na_i Z - 2 N \frac{\p \Omega}{\p Z^*}
{}^{(g)}\na_i N \mid^2\\ \nonumber
&-& \frac{1}{16} N^2 \mid Z {}^{(g)}\na_i Z^* - Z^* {}^{(g)} \na_i Z \mid^2.
\een
It turns out that the relation between electric and magnetic potentials in the static spacetime causes that the last term in (\ref{ricci})) disappear
and one can conclude that $(\Sigma, ~\Omega^2 g_{ij}) $ is an asymptotically flat complete three-dimensional manifold with non-negative scalar curvature
and vanishing mass. Next, the implementation of positive energy theorem implies that the manifold in question is isometric to $(R^3,~\delta_{ij})$.

The requirements for the positive energy theorem point out that it cannot be implemented for $(\Sigma_+,~\Omega_+^2 g_{ij})$ \cite{bun87}. However
they are satisfied for 
$$(\Sigma, ~g_{ij}) = (\Sigma_+,~\Omega_+^2 g_{ij}) \cup (\Sigma_- \cup \{p\},~\Omega_-^2 g_{ij}), $$
where $\{p\}$ is a point at infinity at $\Sigma_-$ \cite{bun87, mas92}. On the other hand,
the conformal flatness of $(\Sigma, ~g_{ij})$ entails its spherical symmetry \cite{bun87, heu96}.

 The arguments, presented for instance in \cite{heu96,bun87,mas92,heu94}, lead to the final conclusion that the metric $g_{ij}$ is spherically symmetric
and we arrive at the uniqueness of {\it photon sphere} characterized by ADM mass $M$ and electric and magnetic charges, as the only static spherically symmetric 
spacetime, possessing {\it photon sphere},
in
Einstein-Maxwell gravity with electric and magnetic potentials.\\

Summing it all up, we achieve the main result, the uniqueness of {\it photon sphere} for non-extremal Reissner-Nordstr\"om electric-magnetic system.\\
\noindent
{\bf Theorem}:\\
Suppose that the set $(M^3,~g_{ij},~N,~\psi_{F}, ~\psi_{B})$ is the asymptotic to the static non-extremal Einstein-Maxwell
black hole spacetime with electric and magnetic charges. Moreover the spacetime in question has
{\it photon sphere}
$(P^3,~h_{ij}) 
\hookrightarrow (R \times M^3,~ -N^2 dt^2 + g_{ij}dx^i dx^j),$
which can be regarded as the inner boundary of $R \times M^3$. 
Suppose further, that $M$,~ $Q_{(F)}$ and 
$Q_{(B)}$ are the ADM mass and the total charges
connected with Maxwell electric and magnetic and fields 
 of $(R \times M^3
 ,~ -N^2 dt^2 + g_{ij}dx^i dx^j)$.
Then, $(R \times M^3
,~ -N^2 dt^2 + g_{ij}dx^i dx^j)$ 
is isometric to the region 
exterior to the {\it photon sphere} 
in the electrically and magnetically charged  non-extremal Einstein-Maxwell black hole spacetime. 

\section{Conclusions}
Our paper is devoted to the problem of uniqueness of black hole {\it photon sphere} in Einstein-Maxwell gravity with electric and magnetic charges.
Having in mind the special features of electric and magnetic fields in the spacetime with
asymptotically timelike Killing vector field, which is orthogonal to the hypersurface of constant time, we find the functional dependence among lapse function 
and electric, magnetic potentials. It authorizes that the Ricci curvature scalar of {\it photon sphere} is constant scalar curvature one.

The conformal positive energy and positive energy theorems allow us to find the two alternative proofs of the uniqueness of a static, non-extremal asymptotically flat black hole 
{\it photon sphere} in Einstein-Maxwell gravity with electric and magnetic charges (Reissner-Nordstr\"om electric magnetic black hole {\it photon sphere}).

\acknowledgments 
MR was partially supported by Grant No. 2022/45/B/ST2/00013 of the National Science Center, Poland.






\begin{thebibliography}{99}

%
\def\cmp#1#2#3#4{\emph{#4}, \emph{ Commun. Math. Phys.} {\bf #1} #2 (#3)}
\def\lmp#1#2#3#4{\emph{#4}, \emph{ Lett. Math. Phys.} {\bf #1} #2 (#3) }
\def\hpa#1#2#3#4{\emph{#4}, \emph{ Hell. Phys. Acta} {\bf #1} #2 (#3) }
\def\grg#1#2#3#4{\emph{#4}, \emph{ Gen. Rel. Grav.} {\bf #1} #2 (#3) }


\def\pr#1#2#3#4{\emph{#4}, \emph{ Phys. Rev.} {\bf #1} #2 (#3)}
\def\prl#1#2#3#4{\emph{#4}, \emph{ Phys. Rev. Lett.} {\bf #1} #2 (#3)}
\def\prd#1#2#3#4{\emph{#4}, \emph{ Phys. Rev. D} {\bf #1} #2 (#3)}
\def\pl#1#2#3#4{\emph{#4}, \emph{ Phys. Lett.} {\bf #1} #2 (#3) }
\def\pla#1#2#3#4{\emph{#4}, \emph{ Phys. Lett. A} {\bf #1} #2 (#3)}
\def\plb#1#2#3#4{\emph{#4}, \emph{ Phys. Lett. B} {\bf #1} #2 (#3)}
\def\prep#1#2#3#4{\emph{#4}, \emph{ Phys. Reports} {\bf #1} #2 (#3) }
\def\phys#1#2#3#4{\emph{#4}, \emph{ Physica} {\bf #1} #2  (#3) }
\def\jcp#1#2#3#4{\emph{#4}, \emph{ J. Comput. Phys.} {\bf #1} #2 (#3) }
\def\jmp#1#2#3#4{\emph{#4}, \emph{ J. Math. Phys.} {\bf #1} #2 (#3) }
\def\jpm#1#2#3#4{\emph{#4}, \emph{ J. Phys. A: Math. Gen.} {\bf #1} #2 (#3) }
\def\cpr#1#2#3#4{\emph{#4}, \emph{ Computer Phys. Rept.} {\bf #1} #2 (#3) }
\def\cqg#1#2#3#4{\emph{#4}, \emph{ Class. Quant. Grav.} {\bf #1} #2 (#3)}
\def\cma#1#2#3#4{\emph{#4}, \emph{ Computers Math. Applic.} {\bf #1} #2  (#3)}
\def\mc#1#2#3#4{\emph{#4}, \emph{ Math. Compt.} {\bf #1} #2 (#3)}
\def\apj#1#2#3#4{\emph{#4}, \emph{ Astrophys. J.} {\bf #1} #2 (#3) }
\def\apjs#1#2#3#4{\emph{#4}, \emph{ Astrophys. J. Suppl.} {\bf #1} #2 (#3)}
\def\apjl#1#2#3#4{\emph{#4}, \emph{ Astrophys. J. Lett.} {\bf #1} #2 (#3) }
\def\acta#1#2#3#4{\emph{#4}, \emph{ Acta Astronomica} {\bf #1} #2 (#3) }
\def\apl#1#2#3#4{\emph{#4}, \emph{ Ann. Physik. (Leipzig)} {\bf #1} #2 (#3) }
\def\amjp#1#2#3#4{\emph{#4}, \emph{Am. J. Phys.} {\bf #1} (#3) #2}
\def\anp#1#2#3#4{\emph{#4}, \emph{ Ann. Phys.} {\bf #1} (#3) #2}
\def\sa#1#2#3#4{\emph{#4}, \emph{ Sov. Astro.} {\bf #1} (#3) #2}
\def\sia#1#2#3#4{\emph{#4}, \emph{ SIAM J. Sci. Statist. Comput.} {\bf #1} (#3) #2}
\def\aa#1#2#3#4{\emph{#4}, \emph{ Astron. Astrophys.} {\bf #1} #2 (#3)}
\def\mnras#1#2#3#4{\emph{#4}, \emph{ Mon. Not. R. Astr. Soc.} {\bf #1} (#3) #2}
\def\npb#1#2#3#4{\emph{#4}, \emph{ Nucl. Phys. B} {\bf #1} (#3) #2}
\def\prsla#1#2#3#4{\emph{#4}, \emph{ Proc. R. Soc. London, Ser. A} {\bf #1} (#3) #2}
\def\jhep#1#2#3#4{\emph{#4}, \emph{ JHEP} {\bf #1} (#2) #3}
\def\jcap#1#2#3#4{\emph{#4}, \emph{ J. Cosmol. Astropart. Phys.} {\bf #1} #2 (#3)}
\def\nuc#1#2#3#4{\emph{#4}, \emph{ Nuovo Cimento B } {\bf #1} (#3) #2}
\def\ijmp#1#2#3#4{\emph{#4}, \emph{ Int. J. Mod. Phys. D} {\bf #1} (#3) #2}
\def\atmp#1#2#3#4{\emph{#4}, \emph{ Adv. Theor. Math. Phys.} {\bf #1} #2 (#3) }
\def\ptps#1#2#3#4{\emph{#4}, \emph{ Prog. Theor. Phys. Suppl.} {\bf #1} #2 (#3)}
\def\lmp#1#2#3#4{\emph{#4}, \emph{ Lett. Math. Phys.} {\bf #1} #2 (#3)}
\def\cpam#1#2#3#4{\emph{#4}, \emph{ Comm. Pure Appl. Math.}  {\bf #1} #2 (#3) }
\def\adv#1#2#3#4{\emph{#4}, \emph{ Adv. Phys.}  {\bf #1} (#3) #2}
\def\zh#1#2#3#4{\emph{#4}, \emph{ Zh. Eksp. Teor. Fiz.}  {\bf #1} (#3) #2}

\def\jams#1#2#3#4{\emph{#4}, \emph{ J. Austral. Math. Soc. B} {\bf #1} (#3) #2}
\def\appa#1#2#3#4{\emph{#4}, \emph{ Acta Phys. Polonica A} {\bf #1}, (#3) #2}
\def\nat#1#2#3#4{\emph{#4}, \emph{Nature} {\bf #1}, (#3) #2}
\def\science#1#2#3#4{\emph{#4}, \emph{Science} {\bf #1}, (#3) #2}
\def\arcmp#1#2#3#4{\emph{#4}, \emph{Annual Rev. of Cond. Matter Physics} {\bf #1}, (#3) #2}
\def\jcap#1#2#3#4{\emph{#4}, \emph{JCAP} {\bf #1}, (#3) #2}
\def\conphy#1#2#3#4{\emph{#4}, \emph{Contemporary Physics} {\bf #1}, (#3) #2}
\def\ptps#1#2#3#4{\emph{#4}, \emph{ Prog. Theor. Phys. Suppl.} {\bf #1} #2 (#3) }
\def\ptp#1#2#3#4{\emph{#4}, \emph{ Prog. Theor. Phys.} {\bf #1} #2 (#3)}
\def\ptpexp#1#2#3#4{\emph{#4}, \emph{ Prog. Theor. Exp. Phys.} {\bf #1} #2 (#3)}

\def\apjsup#1#2#3#4{\emph{#4}, \emph{ Astrophys. J. Suppl. Ser.} {\bf #1} (#3) #2}
\def\mplb#1#2#3#4{\emph{#4}, \emph{ Mod. Phys. Lett. B} {\bf #1} (#3) #2}
\def\ijmpd#1#2#3#4{\emph{#4}, \emph{ Int. J. Mod. Phys. D} {\bf #1} (#3) #2}

\def\gravcos#1#2#3#4{\emph{#4}, \emph{ Grav. Cosmol.} {\bf #1} #2 (#3)}
\def\ajp#1#2#3#4{\emph{#4}, \emph{ Am. J. Phys.} {\bf #1} (#3) #2}
\def\appb#1#2#3#4{\emph{#4}, \emph{ Acta Phys. Polon. B} {\bf #1} (#3) #2}
%
\def\hepph#1#2{{ hep-ph }{#1} (#2)}
\def\hepth#1#2{{ hep-th }{#1} (#2)}
\def\arx#1#2{{ arXiv~}{#1} (#2)}
\def\astroph#1#2{{ astro-ph }{#1} (#2)}
\def\grqc#1#2{{ gr-qc }{#1} (#2)}
\def\ibid#1#2#3#4{\emph{#4}, {\it ibid.} {\bf #1} #2 (#3)}
\def\cag#1#2#3#4{\emph{#4}, \emph{ Commun. Anal. Geom.} {\bf #1} #2 (#3) }
\def\contmath#1#2#3#4{\emph{#4}, \emph{ Contemp. Math.} {\bf #1} #2 (#3)}
\def\epjc#1#2#3#4{\emph{#4}, \emph{ Eur. Phys. J. C} {\bf #1} #2 (#3) }
\def\revphys#1#2#3#4{\emph{#4}, \emph{Reviews in Phys.} {\bf #1} #2 (#3) }
\def\science#1#2#3#4{\emph{#4}, \emph{ Science} {\bf #1} #2 (#3) }

%
\bibitem{eht1}
K. Akiyama et al., \apjl{875}{L1}{2019}{First M87 Event Horizon Telescope Results. I. The Shadow of the Supermassive Black Hole}.
\bibitem{eht2}
K. Akiyama et al., \apjl{930}{L12}{2022}{First Sagitarius A* Event Horizon Telescope Results.1. The shadow of the supermassive black hole in the center of the Milky Way}.

\bibitem{eht mag1}
K. Akiyama et al., \apjl{910}{L12}{2021}{First M87 Event Horizon Telescope Results. VII. Polarization of the Ring}.
\bibitem{eht mag2}
K. Akiyama et al., \apjl{910}{L13}{2021}{First M87 Event Horizon Telescope Results. VIII. Magnetic Field Structure near The Event Horizon}.
\bibitem{eht mag3}
K. Akiyama et al., \apjl{957}{L20}{2023}{First M87 Event Horizon Telescope Results. IX. Detection of Near-horizon Circular Polarization}.

\bibitem{vir00}
K.S. Virbhadra and G.F.R. Ellis, \prd{62}{084003}{2000}{Schwarzschild black hole lensing}.
\bibitem{cla01}
C-M. Claudel, K.S. Virbhadra, and G.F.R. Ellis, \jmp{42}{818}{2001}{The geometry of photon surfaces}.


\bibitem{heu96}
M. Heusler, {\it Black Hole Uniqueness Theorems}, 
Cambridge: Cambridge University Press, 1996.

\bibitem{ced15a}
C. Cederbaum, \contmath{667}{86}{2017}{Uniqueness of photon spheres in static vacuum asymptotically flat spacetimes}.

\bibitem{yaz15}
S. Yazadjiev, \prd{91}{123013}{2015}{Uniqueness of the static spacetimes with a photon sphere in Einstein-scalar field theory}.
\bibitem{yaz15b}
S. Yazadjiev and B. Lazov, \cqg{32}{165021}{2015}{Uniqueness of the static Einstein-Maxwell spacetimes with a photon sphere}.


\bibitem{ced15}
C. Cederbaum and G. Galloway, \cag{25}{303}{2017}{Uniqueness of photon spheres via positive mass rigidity}.
\bibitem{ced16}
C. Cederbaum and G. Galloway, \cqg{33}{075006}{2016}{Uniqueness of photon spheres in electro-vacuum spacetimes}.

\bibitem{rog16}
M. Rogatko, \prd{93}{064003}{2016}{Uniqueness of photon sphere for Einstein-Maxwell-dilaton black holes with arbitrary coupling constant}.

\bibitem{yaz16}
S. Yazadjiev and B. Lazov, \prd{93}{064003}{2016}{Classification of the static and asymptotically flat Einstein-Maxwell-dilaton spacetimes with a photon sphere}.

\bibitem{tom17}
Y. Tomikawa, T. Shiromizu, and K. Izumi, \ptpexp{2017}{033E03}{2017}{On the uniqueness of the static black hole with conformal scalar hair}.
\bibitem{tom17b}
Y. Tomikawa, T. Shiromizu, and K. Izumi, \cqg{34}{15504}{2017}{On uniqueness of static spacetimes with non-trivial conformal scalar field}.
\bibitem{yos17}
H. Yoshino, \prd{95}{044047}{2017}{Uniqueness of static photon surfaces: Perturbative approach}.

\bibitem{kog20}
Y. Koga, \prd{101}{104022}{2020}{Photon surfaces as pure tension shells: Uniqueness of thin shell wormholes}.
\bibitem{yaz21}
S. Yazadjiev, \prd{104}{124070}{2021}{Classification of static asymptotically flat spacetimes with a photon sphere in Einstein-multiple-scalar field theory}.

\bibitem{jah19}
S. Jahns, \cqg{36}{235019}{2019}{Photon sphere uniqueness in higher-dimensional electrovacuum spacetimes}.
\bibitem{bud20}
M. Budgen, \cqg{37}{015001}{2020}{Trapped photons in Schwarzschild-Tangherlini spacetimes}
\bibitem{ced21}
C. Cederbaum and G. Galloway, \jmp{62}{032504}{2021}{Photon surfaces with equipotential time slices}.





\bibitem{shi17}
T. Shiromizu, Y. Tomikawa, K. Izumi, and H. Yoshino, \ptpexp{2017}{033E01}{2017}{Area bound for surface in a strong gravity}.
\bibitem{yan20}
R. Q. Yang and H. Lu, \epjc{80}{949}{2020}{Universal bounds on the size of a black hole}.
\bibitem{fen20}
X. H. Feng and H. Lu, \epjc{80}{551}{2020}{On the size of rotating black holes}.



\bibitem{gib16}
G. W. Gibbons and C. M. Warnick, \plb{763}{169}{2016}{Aspherical photon and anti-photon surfaces}.
\bibitem{yos17b}
H. Yoshino, K. Izumi, T. Shiromizu, and Y. Tomikawa, \ptpexp{2017}{063E0}{2017}{Extension of photon surfaces and their area: Static and stationary spacetimes}.
\bibitem{sho17}
A. Shoom, \prd{96}{084056}{2017}{Metamorphoses of a photon sphere}.
\bibitem{gal19}
D. V. Gal'tsov and K. V. Kobialko, \prd{99}{084043}{2019}{Completing characterization of photon orbits in Kerr and Kerr-Newman metrics}.

\bibitem{ced19}
C. Cederbaum and S. Jahns, \grg{51}{79}{2019}{Geometry and topology of the Kerr photon region in the phase space}, ~\ibid{51}{154(E)}{2019}{Correction to: Geometry and topology 
of the Kerr photon region in the phase space}.
\bibitem{gal19b}
D. V. Gal'tsov and K. V. Kobialko, \prd{100}{104005}{2019}{Photon trapping in static axially symmetric spacetime}.

\bibitem{cao21}
L. -M. Cao and Y. Song, \epjc{81}{714}{2021}{Quasi-local photon surfaces in general spherically symmetric spacetimes}.

\bibitem{yos20}
H. Yoshino, K. Izumi, T. Shiromizu, and Y. Tomikawa, \ptpexp{2020}{023E02}{2020}{Transversely trapping surfaces: Dynamical version }.
\bibitem{kob20}
K. V. Kobialko and D. V. Gal'tsov, \epjc{80}{527}{2020}{Photon regions and umbilic conditions in stationary axisymmetric spacetimes}.

\bibitem{kog21}
Y. Koga, T. Igata, and K. Nakashi, \prd{103}{044003}{2021}{Photon surfaces in less symmetric spacetimes}.


\bibitem{kob21}
K. V. Kobialko, I. Bogush, and D. V. Gal'tsov, \prd{104}{044009}{2021}{Killing tensors and photon surfaces in foliated spacetimes}.
\bibitem{kob22}
K. V. Kobialko, I. Bogush, and D. V.  Gal'tsov, \prd{106}{024006}{2022}{Slice-reducible conformal Killing tensors, photon surfaces, and shadows}.


\bibitem{kob22a}
K. V. Kobialko, I. Bogush, and D. V.  Gal'tsov, \arx{2208.02690}{2022}
{\it The geometry of massive particle surfaces}. 
\bibitem{bog23}
I. Bogush,K. V. Kobialko, and D. V.  Gal'tsov, \arx{2306.12888}{2023}
{\it Glued massive particle surfaces}.



\bibitem{dal18}
Y. Dallilar et al. \science{358}{1299}{2017}{A precise measurement of the magnetic field in the corona of the black hole binary V404 Cygni}.


\bibitem{isr67}
W. Israel, \cmp{8}{245}{1967}{Event horizons in static electrovac space-time}.




\bibitem{sim99}
W. Simon, \lmp{50}{275}{1999}{Conformal Positive Mass Theorems}.



\bibitem{mar02}
M. Mars and W. Simon, \atmp{6}{279}{2002}{On uniqueness of static Einstein-Maxwell-dilaton black holes}.
\bibitem{gib02}
G.W. Gibbons, D. Ida, and T. Shiromizu, \prl{89}{041101}{2002}{Uniqueness and nonuniquess of static black holes in higher dimensions}.
\bibitem{gib02a}
G.W. Gibbons, D. Ida, and T. Shiromizu, \prd{66}{044010}{2002}{Uniqueness of (dilatonic) charged black holes and black p-branes in higher dimensions}.
\bibitem{rog02}
M. Rogatko, \cqg{19}{L151}{2002}{Uniqueness theorem for static black hole solutions of $\sigma$-models in higher dimensions}.
\bibitem{rog03}
M. Rogatko, \prd{67}{084025}{2003}{Uniqueness theorem of static degenerate and nondegenerate charged black holes in higher dimensions}.
\bibitem{rog04}
M. Rogatko, \prd{70}{044023}{2004}{Uniqueness theorem for generalized Maxwell electric and magnetic black holes in higher dimensions}.
 \bibitem{rog06}
 M. Rogatko, \prd{73}{124027}{2006}{Classification of static charged black holes in higher dimensions}.
\bibitem{rog13}
M. Rogatko, \prd{88}{024051}{2013}{Uniqueness of charged static asymptotically flat black holes in dynamical Chern-Simons gravity}.
\bibitem{rog22}
M. Rogatko, \prd{105}{104021}{2022}{Classification of static black holes in Einstein phantom-dilaton Maxwell--anti-Maxwell gravity systems}.
\bibitem{wormholes18}
M. Rogatko, \prd{97}{024001}{2018}{Uniqueness of higher-dimensional phantom field wormholes}, M. Rogatko, \ibid{97}{064023}{2018}{Uniqueness of higher-dimensional 
Einstein-Maxwell-phantom dilaton field wormholes}.




\bibitem{isr67}
W. Israel, \cmp{8}{245}{1967}{Event horizons in static electrovac space-time}.


\bibitem{hoe76}
C. Hoenselaers, \ptp{55}{46}{1976}{Multipole moments of electrostatic space-times}.
\bibitem{sim92}
W. Simon, \cqg{9}{241}{1992}{Radiative Einstein-Maxwell spacetimes and 'no-hair' theorems}.

\bibitem{posen}
R. Schoen and S.T. Yau, \cmp{65}{45}{1979}{On the proof of the positive mass conjecture in general relativity},\\
E. Witten, \cmp{80}{381}{1981}{A new proof of the positive energy theorem}.
\bibitem{heu94}
M. Heusler, \cqg{11}{L49}{1994}{On the uniqueness of Reissner-Nordstr\"om solution with electric and magnetic charge}.
\bibitem{bun87}
G. L. Bunting and A. K. M. Masood-ul-Alam, \grg{19}{147}{1986}{Nonexistence of multiple black holes in asymptotically Euclidean static vacuum space-time}.
\bibitem{mas92}
A. K. M. Masood-ul-Alam, \cqg{9}{L53}{1992}{Uniqueness proof of static charged black hole revisited}.





\end{thebibliography}
\end{document}